\documentstyle{mn}
\newcommand{\be}{\begin{equation}}
\newcommand{\ee}{\end{equation}}
\newcommand{\cd}{$\cdots$}
\title{Stellar Forensics: II - Millisecond Pulsar Binaries}
\author[B. M. S. Hansen \& E. S. Phinney]
{Brad M. S. Hansen$^{1,2}$\thanks{email:{\bf hansen@cita.utoronto.ca}}
 \& E. Sterl Phinney$^2$ \\
$^1$ Canadian Institute for Theoretical Astrophysics, University of
Toronto, Toronto, ON M5S 3H8, Canada \\
$^2$ Theoretical Astrophysics, 130-33, California Institute of Technology,
Pasadena, CA 91125}
\date{28 August 1997}

\pagerange{\pageref{firstpage}--\pageref{lastpage}}
\pubyear{1997}
\begin{document}

\maketitle
\label{firstpage}

\begin{abstract}
We use the grid of models described in paper~I to analyse those millisecond
pulsar binaries whose secondaries have been studied optically. In particular,
we find cooling ages for these binary systems that range from $< 1 \, \rm Gyr$
to $\sim \rm 15 \, Gyr$. Comparison of cooling ages and characteristic spin down
ages allows us to constrain the initial spin periods and spin-up histories
for individual systems, showing that at least some millisecond pulsars had
sub-Eddington accretion rates and long magnetic field decay times.
\end{abstract}


\section{Introduction}
The ages of millisecond pulsars are important for understanding both their
nature and origin. Ages are usually estimated from the characteristic 
age $ t_P = P/2 \dot{P}$, but such estimates could be seriously in error if the current spin
period is still close to the initial spin period at the beginning of the
millisecond pulsar phase. If the average millisecond pulsar is significantly younger than
its characteristic age, as suggested by Lorimer et al. (1995a),
 then it would affect current ideas about magnetic field decay in
such stars (Kulkarni~1986; Camilo, Thorsett \& Kulkarni~1994) as well as the possible birth-rate discrepancy
between low mass binary pulsars and low mass X-Ray binaries 
(Kulkarni \& Narayan~1988; van den Heuvel~1995). 

Many millisecond pulsars are found in binaries, often with low mass degenerate
companions. Accurate modelling of the cooling of the companions can allow one
to estimate the age of the system (or rather the age of this particular
incarnation). Since millisecond pulsars are thought to be spun-up as the result
of accretion from the companion star, the pulsar will begin to spin down
at approximately the same time as the companion shrinks within its Roche lobe, ending mass loss
and beginning the process of cooling to its final degenerate white dwarf configuration.
Hence,
the cooling age of the white dwarf should represent the millisecond pulsar age as
well.

In paper~I, 
 we calculated accurate cooling models for
the low mass helium core white dwarfs thought to be the companions in these low mass
binary pulsar systems. In this paper we shall now apply the models to the 
optical observations of these systems to infer the cooling ages for such objects.
Section~\ref{Pulsar} reviews the basic concepts used to infer the age of a pulsar
from its spin parameters as well as simple models of magnetic field decay in
such a pulsar. In section~\ref{Results} we apply our cooling models to the
observational data and in section~\ref{Discuss} we discuss the implications.

\section{Pulsar Spin Down}
\label{Pulsar}
For a pulsar with period $P$ and  spin-down rate $\dot{P}$ related by $ \dot{P} \propto P^{2-n}$, the
age is given by
\be
 t = \frac{P}{(n-1)\dot{P}} \left( 1  - \left( \frac{P_0}{P}\right)^{n-1} \right),
\label{spin-down}
\ee
where $n$ is the braking index and $ P_0$ is the initial period. For magnetic dipole radiation
$n=3$, leading to the familiar expression for the pulsar characteristic age $t_P=P/2\dot{P}$.
For some young pulsars, measurements yield a range for $n$ of 2.0 to 2.8 ( Lyne 1996 and references
therein). However, $n$ is not known for any millisecond pulsars.
 Because characteristic ages for millisecond pulsars are of the order of Gyrs or higher,
 there is a
very real danger that estimates of the pulsar age based on this quantity will be gross overestimates
because of the second factor in equation~(\ref{spin-down}).

The characteristic time is calculated from the spin-down rate $\dot{P}$, which can also be seriously
affected by the proper motion of the pulsar (e.g. Camilo et al. 1994).
 This is because of the changing doppler shift 
 (Shklovskii 1970), which produces an additional contribution to the period
derivative such that 
\begin{eqnarray} \frac{\dot{P}}{P} & = & \left(\frac{\dot{P}}{P}\right)_{\rm i} + \frac{V_{\rm \perp}^2}{c D}  \\
& = & \left(\frac{\dot{P}}{P}\right)_{\rm i}+ 2.432 \times 10^{-21} \left(\frac{\mu}{\rm mas.yr^{-1}}
\right)^2
 \left(\frac{D}{\rm kpc}\right), \nonumber \end{eqnarray}
where 
$\left(\dot{P}/P\right)_{\rm i} $ is the intrinsic contribution due to pulsar rotation, $D$ is the
distance to the pulsar, $V_{\perp} \propto D \mu$ is the transverse velocity and $\mu$ is the
proper motion of the pulsar.
Once $\mu$
 is measured, we may remove this contribution to obtain the intrinsic characteristic time.
Let us define $t_{P}^*$ as the characteristic age before the correction and $t_{P}$ as the
intrinsic pulsar characteristic age.
Thus, with $n=3$, the intrinsic characteristic age is found by
\be
t_P = \frac{t_P^*}{1 - 2 V_{\perp}^2 t_P^*/c D}.
\ee
 We may then obtain an expression for the ratio of the true age of the pulsar to 
 $t^*_{P}$ by
\be
\frac{t}{t^*_P} = \left( 1 - \left(\frac{P_0}{P}\right)^2\right)/\left(1 - \frac{2 V_{\perp}^2 t^*_P}
{c D}\right).
\label{ttp}
\ee
Apart from the initial period $P_0$, all the quantities on the right side are measurable in
principle. An independent age estimate, such as that from the cooling of a white dwarf,
can thus be used to estimate the initial period. If the companion is undetectable then
only a lower limit to the age is possible, which in turn translates into an upper limit
on $P_0$ (as long as $2 V_{\perp}^2 t_P^*/c D < 1$, which is the case unless the intrinsic
period derivative is negative).

There is an additional complication, namely the possibility of magnetic field decay.
For magnetic dipole radiation, the spin down rate is given by
\be
\dot{P} = \left( \frac{8 \pi^2 R^6}{3 I c^3} \right) \frac{B^2}{P},
\ee
where $B$ is the magnetic field strength on the surface of the neutron star
($B$ is in fact  $B \sin \alpha$, where $\alpha$ is the angle between 
the magnetic dipole axis and the spin axis, and so `decay' could be the result of field alignment).
 Thus, decay of $B$ will
alter the spin-down rate. The possibility of field decay was first advanced
 by Gunn \& Ostriker (1970)
but its occurrence is still not conclusively proven. Narayan \& Ostriker (1991) have argued for an exponential
field decay with a decay time of $\sim 10^7$ years. Bhattacharya et al. (1992), on the other hand, find
no evidence for field decay. Furthermore, there may be a subdivision between `normal' pulsars
and `recycled' pulsars (i.e., those spun up through accretion in binaries). Kulkarni (1986),
Verbunt, Wijers \& Burm (1990) and Camilo et
al (1994) have argued that the fields on recycled pulsars do not decay, based on the ubiquitous nature
of the field strength $\rm \sim 10^{8-9} G$ amongst this population of objects and on the presence of cool
companions such as the ones we address here.

 Thus, we shall also consider magnetic field decay of the
form
$ B = B_0 \, exp(-t/t_{\rm D}) $, where $ t_{\rm D}$ is the decay time. In this case, the equivalent of 
equation~(\ref{ttp}) is
\be \frac{t}{t_{\rm D}} = \frac{1}{2} \ln \left[ 1  + \frac{2 t_P^*}{t_{\rm D}} 
 \frac{\left( 1 - \left( P_0/P\right)^2 \right)}{\left( 1 - 2 V_{\perp}^2
t_P^*/c D \right)}
 \right].
\label{decay1}
\ee

This introduces a second unknown parameter $ t_{\rm D}$ into the equation. Nevertheless, we can obtain a
lower limit to $ t_{\rm D}$ by setting $ P_0=0$, so that, using equation~(\ref{decay1}), we obtain
\be
\frac{t_{\rm D}}{2 t} \left( exp(2 t/t_{\rm D}) - 1 \right) < \frac{t_P^*/t}{ \left( 1 - 2 V_{\perp}^2 t_P^*/c D \right)}=\frac{t_p}{t}.
\label{decay2}
\ee
Once we have an estimate for $t$, the left-hand side is a monotonically decreasing
function of $ t_{\rm D}$ and the right-hand side consists of measurable quantities. Hence, we may
constrain the exponential field decay time.

\section{Results}
\label{Results}

Most low mass binary pulsars have helium core companions (inferred from their mass since
one requires $\sim 0.5 M_{\odot}$ to burn helium to carbon in a stellar interior),
although three have companions whose masses indicate that they are carbon/oxygen
white dwarfs. These measurements come from pulsar timing and are shown in Table~\ref{TimID}
for those pulsars discussed here. In order to determine
a cooling age, we need optical identifications of the companions. Those
systems with detections or upper limits are shown in Table~\ref{OptID}. It
is these systems that we shall now discuss in turn. We note that, while the following 
analysis is quantitative, such an exercise is likely to contain small residual systematic uncertainties
resulting both from the use of a heterogeneous sample set from a number of independent
groups using different instruments and analysis procedures as well as uncertainties in
the atmospheric composition and modelling. The cooling curves used
in this procedure will be made public to enable improved estimates to be made in the future with
better understanding of individual measurement errors and better atmospheric models.

To determine a conservative and robust effective temperature range from the observed colours,
 we have assumed 
a black body spectrum. We use the bandpasses as
described in Bessell (1979).
In the case of the more massive white dwarfs, we use the pure hydrogen broad band colours of 
Bergeron, Saumon \& Wesemael (1995) (their calculations do not extend to low enough log~$g$). Figure~\ref{tc} shows the black-body relation between
temperature and $V-I$ colour (the most common colour amongst this sample). For $V-I<1.2$/$T_{\rm eff}>$~4500~K,
 the black body colour is in good agreement with the Bergeron et al colours. At the 
lower temperature end, the black body approximation is expected to fail because
of the contribution of collisionally induced absorption in the hydrogen atmospheres. In Figure~\ref{tc}
we use the log~$g$=7.5 curves to give us a limit on the deviation of our black-body colours from
the true colours. We do this because the collisionally induced absorption is an inherently high
density phenomenon (since it requires collisional interactions) and thus the deviations from
a black body should be less for lower gravities (which have lower photospheric pressures). Thus, by
using this procedure we believe our temperature estimates are robust.


In the following sections, we shall derive cooling ages for the various binary systems, using
the cooling models described in paper~I. When the companion is thought to be a carbon/oxygen
core white dwarf as opposed to a helium core white dwarf, we use the latest
 models derived using the model of Wood (1992). 
We shall also convert these results into constraints on the initial periods of the millisecond
pulsars. In order to obtain conservative estimates, we consider dispersion measure
distances to be accurate to within $30 \%$, transverse velocities up to 100 $\rm
km.s^{-1}$ for systems without a measured proper motion,\footnote{Millisecond pulsar
velocities are lower than the average velocities of younger pulsars (Lyne \& Lorimer 1994, Hansen
\& Phinney 1997)}  and a braking index between $n=2$
and $n=3$.

\subsection{PSR J0437-4715}
\label{J0437}
We shall start with the best determined case. The PSR J0437-4715 system contains the closest known
millisecond pulsar, at a nominal dispersion measure distance of 0.14 kpc, with a 30\% error 
 (Taylor
\& Cordes 1993). Bell et al (1996) report a lower limit of the parallax distance (from pulsar timing)
of 90 pc.
 Sandhu et al (1997), using improved dispersion measurements,
 report a timing parallax distance of $0.178 \pm 0.026$ kpc.
VLBI parallax measurements are also expected, which will eventually determine the most accurate
distance to this pulsar.
PSR~J0437-4715
 has a large proper motion ($\mu = 135\pm 4 \, \rm mas \, yr^{-1}$)
(Bell et al. 1995), which gives
it a transverse velocity of 115 $\rm km \, s^{-1}$ at 180 pc. This affects the spin-down rate, leading to
a kinematically corrected characteristic age of 4.5-50.3 Gyr, depending on the distance to the
pulsar based on the parallax. 

Optical observations of the companion have been carried out by 
Bell, Bailes \& Bessell (1993),
Bailyn (1993) and Danziger, Baade \& Della Valle (1993). Between them these three groups measured 
$B,V,R$ and $I$
magnitudes. The best data is due to Danziger et al and we will use their measurements.
Using the $V-I$ colour and our black-body approximation,
 we obtain a temperature estimate of $T_{\rm eff} = 4600 \pm 200$ K. 

Given the above parallax distance we can find
the absolute magnitude range allowed, namely
$M_{\rm I}$= 13.2 $\pm$ 0.3.
 Our models must satisfy
both the temperature and magnitude requirements. We can find consistent solutions for 
all masses 0.15-0.375 $\rm M_{\odot}$ models as shown in Figure~\ref{Solve0437} (If we use only
the lower distance limit of Bell et al then solutions for all masses are possible). These models are
for a hydrogen envelope of $3 \times 10^{-4}\rm  M_{\odot}$ (thick hydrogen envelope in the
terminology of paper~I). Similar analysis with thin hydrogen envelope models ($10^{-6} 
\rm M_{\odot}$) yields similar answers, with cooling ages $\sim$ 0.3 Gyr smaller on average.

Further timing measurements by Sandhu et al. have detected, in addition to the parallax,
 a rate of
change in
the projected orbital separation $ x = a \sin i$, which they interpret as a change in the inclination
angle. Assuming that this is the result of the known proper motion, this implies an upper 
limit to the inclination angle ($i < 43^{\circ}$) and  a new lower limit to the mass of the companion, 
$ M \sim 0.22 \, \rm M_{\odot}$. This constrains the mass range to lie between 0.22-0.375 $\rm M_{\odot}$
(for a 1.4 $\rm M_{\odot}$ neutron star)
and the cooling age is thus 2.5-5.3 Gyr. If we reverse the argument, we can use the upper limit on
the companion mass to constrain the neutron star mass to be $< 3.36 \rm M_{\odot}$\footnote{Our
results differ from those Sandhu et al. in the interpretation of the optical measurements. They used
results based on an average of the various different measurements, while we use only what we consider to
be the best measurement, i.e. that of Danziger et al.}.
Furthermore, as we can see from Figure~\ref{md}, this further restricts the distance to
the system. Using the companion mass limit, the allowed cooling models and the apparent magnitude, one can
restrict the distance to $<$~188~pc (which implies that the intrinsic period derivative is
 $\dot{P} > 6.16 \times 10^{-21}$). This allows us to set a better upper limit on the timing age
of 14.9 Gyr. If we assume a pulsar mass of 1.4~$\rm M_{\odot}$, then the upper companion mass limit also
gives us a lower limit on the inclination $i > 24^{\circ}$. These various constraints are demonstrated
in Figure~\ref{md}, which illustrates the relationships between the various constraints 
on distance and mass.

If we assume we know the braking index of the pulsar (the default assumption is $n=3$, the value for
magnetic dipole radiation), then we may compare the lowest
allowed cooling age with the largest characteristic age to get an upper limit on the initial spin period. For
pulsar J0437-4715, the proximity and large proper motion mean that the Shklovskii correction
is very important for this comparison. For each possible value of the distance, this
distance-dependent term yields a value for the characteristic age. Similarly, each different distance
leads to a different inferred absolute magnitude. Thus, eliminating the distance,
$ t_P$ may be plotted as a function of
$ M_I$ in Figure~\ref{Solve0437}.
Using the distance and cooling limits derived above, we 
 obtain an upper limit on the initial period of $ P_0 < 0.91 P$ or $ P_0 <
$ 5.3~ms for a braking index $n=3$. If we use $n=2$, we may constrain the initial period to lie in the range
2.4~ms$<P_0<$ 5.3~ms (the lower limit on the timing age is now larger than the upper limit on the
cooling age, thereby allowing us to infer a lower limit for the initial spin period).

For completeness we have also compared the observations with the carbon/oxygen models of
Wood (1992). For J0437-4715, we find no C/O models that are consistent with the observations.
If one lessens the distance restrictions and uses the dispersion measure distance, then one
finds 
that consistent solutions may be obtained for C/O
models with a mass fraction $\sim 10^{-4}$ of surface hydrogen and $M = 0.5 - 0.55 \rm M_{\odot}$.
The cooling ages allowed by this are 5.6-5.9 Gyr, although again it requires a braking
index $\sim 2$.

\subsection{PSR J701012+5307}
\label{J1012}
Even without the benefit of a detailed model, the companion of this system
 is obviously much brighter than it should be if the
system were truly as old as the pulsar characteristic age of $>$5.4 Gyrs 
(Lorimer et al. 1995a). The characteristic age could be even larger if the pulsar
has a significant transverse velocity. 

The system is also closer than average at a distance of 0.52 kpc, thus making it an
ideal candidate for more detailed study. van~Kerkwijk, Bergeron and Kulkarni (1996) have determined
the effective temperature and gravity using the spectroscopic analysis procedure of 
Bergeron, Wesemael \& Fontaine (1991).
The star has $\log~g = 6.75 \pm 0.07$ and $T_{\rm eff}$ = 8550 $\pm$ 25 K
\footnote{As yet unpublished results of Callanan \& Koester (1997) infer a different
gravity of $\log g = 6.4 \pm 0.2$. We will consider both values in our analysis.}. Furthermore, we
can use the photometry of Lorimer et al. (1995a) to infer an absolute magnitude. Combining
these three data in Figure~\ref{Solve1012} we find that we can obtain consistent
solutions for both thick and thin hydrogen atmospheres.

Using our mass-radius relations from paper~I, we can determine that the mass limits on
the companion are 0.157-0.215 $\rm M_{\odot}$ for a thick hydrogen envelope and
0.13-0.183 $\rm M_{\odot}$ for a thin hydrogen envelope. The corresponding cooling
ages are thus 0.18-0.4 Gyr and $\leq$ 0.23 Gyr respectively. Thus, our combined result
for the PSR J1012+5307 companion is that $M \sim 0.165 - 0.215 \, \rm M_{\odot}$ and
the age is  $<0.4$ Gyr.
 

van~Kerkwijk, Bergeron \& Kulkarni (1996) have also measured the radial velocity of the companion, making this a
double lined spectroscopic binary. This allows us to calculate the mass ratio of the two components.
The original published value was 
 $M_{NS}/M_{WD} = 13.3 \pm 0.7$, but it has recently been revised to $M_{NS}/M_{WD} = 10$ (van Kerkwijk,
pers. comm.). Thus, given the companion mass range inferred from the spectroscopic gravity, one
can infer the neutron star mass directly. Using the two different gravities inferred for this object,
we infer neutron star masses of $1.3 - 1.8 M_{\odot}$ (Callanan \& Koester) or $1.7 - 2.1 M_{\odot}$
(van Kerkwijk et al). The former number is in excellent agreement with the original author's estimates.
These estimates are shown in
 Figure~\ref{mass}.


Alberts et al. (1996) claim that a 0.185 $\rm M_{\odot}$ white dwarf burning
 $\sim 2 \times 10^{-3} \rm M_{\odot}$
of hydrogen can have the correct gravity and effective temperature at a cooling age commensurate with
the pulsar's characteristic age (although, as noted above, the characteristic age could increase further). 
Their solar metallicity model does not undergo shell flashes 
 and constitutes this decade's installment
in the conflict between the calculations of Webbink (1975) and Iben and Tutukov (1986)! An 
examination of Figure~1 and Table~1 of Iben and Tutukov (1986)  suggests that the $>50$ yr timesteps
used by Alberts et al. may still not be enough to resolve the shell flash behaviour reported in
Iben \& Tutukov. The shell flash results in the burning of much of the surface hydrogen, which leads to the
difference in surface hydrogen masses that is the source of the different cooling ages. Although Alberts
et al. obtain $\log g\sim$6.7 and $ T_{\rm eff} = 8481$ K, we find that our code (which we
believe to contain more accurate input physics at the cool temperatures appropriate to these
studies) yields a lower $\log~g
\sim$ 6.6 for the same parameters at this temperature, too low to be a consistent solution.

One other uncertainty with this measurement is the possible presence of helium in the
white dwarf atmosphere. Bergeron et al. (1991) have shown that a small amount of helium in
a hydrogen dominated atmosphere can mimic the effect of a larger gravity, a result confirmed
by Reid's (1996) comparison of spectroscopic gravities and gravitational redshifts for cool
white dwarfs. However, when combined with the temperature and luminosity constraints above,
our models provide good agreement with pure hydrogen atmospheres. The admixture of trace 
amounts of helium would not affect the cooling significantly on these timescales but would
measureably raise the gravity. Hence we believe these to be hydrogen dominated white dwarfs.

With the present uncertain status of the observations, our models yield the following results
for the J1012+5307 system. The companion mass lies in the range $0.13-0.21 M_{\odot}$, the
neutron star mass in the range $1.3-2.1 M_{\odot}$, and the age is $<$ 0.4 Gyr.
Allowing for a braking index of $n=2-3$, this
results in the constraint $ P_0 > 5.06$ ms.
Given the above uncertainties about 
the gravity measurement, it is useful to calculate the constraints
without using the gravity. If we use only the magnitude and temperature, then 
the white dwarf mass is constrained to lie in the range
0.12 - 0.32 $\rm M_{\odot}$, and the age is $<$ 0.8 Gyr. Thus, even ignoring the
gravity information, we find that this pulsar is still much younger than
the pulsar characteristic age.

\subsection{PSR B0820+02}

The pulsar in this system has the longest spin-period (864.8 ms), largest magnetic field
($10^{11.5}$G) and
shortest characteristic time (0.13 Gyr)(Of course, these are not independent quantities!) of all the low mass binary pulsars.
The companion has been studied optically by  Kulkarni (1986) and Koester,
Chanmugam \& Reimers (1992). We shall use
the Koester et al. apparent magnitude $V$=22.76 $\pm$ 0.05. van~Kerkwijk \& Kulkarni (1995) have also
identified hydrogen balmer lines in the companion spectrum, yielding an effective temperature
 consistent with the photometric temperature estimate of $15250 \pm 1250$ K 
by  Koester et al.
 The signal to noise ratio is not yet good enough to determine a spectroscopic gravity.

The dispersion measure distance to this pulsar is 1.4 $\pm$ 0.4 kpc. This yields an absolute
$V$ magnitude $M_V$ = 12.1 $\pm$ 0.7. This leads to a problem when we try to find consistent
temperature-luminosity solutions for this pulsar. As shown in Figure~\ref{Solve0820}, the
effective temperature and absolute magnitude give a consistent solution for C/O models with
$ M > 0.8 \, \rm M_{\odot}$. However, these solutions require ages longer than the timing age (even for
$n=2$). If one requires than $t<t_P$, then one is forced to the conclusion that
 the true
distance is larger than the dispersion measure distance. If we use the
characteristic age as a constraint
 (an upper limit to the true age), then we can place some
constraints on the acceptable models using only the temperature constraints and infer what
the true distance would have to be for a consistent solution.


If we assume a helium core white dwarf (allowing for $n=3$ and $n=2$), then the mass is limited to the range 0.27-0.42 $
\rm M_{\odot}$, and
the true distance has to be $\sim$ 3.5 - 4.5 kpc! If we assume a carbon core white dwarf, the mass
range is 0.5-1.0 $\rm M_{\odot}$, with a true distance of $\sim$ 2 - 2.8~kpc. This is probably
the more palatable of the options, which would make this the fourth binary pulsar to
contain a `normal' white dwarf, and the first of those to have not undergone significant
inspiral during its evolution. We note that  Koester et al. (1992) also infer a distance range
of 1.7-3 kpc based on a similar analysis (although their final published solution is 1.7-1.9
kpc, based on the supposed allowed error in the dispersion measure).

If we choose to believe the distance limits, then we are faced with the problem that the
white dwarf is older than the pulsar. One possible way of arranging such a situation is
if the white dwarf is formed from the initially more massive star (e.g. $\sim 8 \rm M_{\odot}$)
which conservatively transfers most of its mass to its initially less massive companion
($\sim 4 \rm M_{\odot}$) thereby creating a system of white dwarf ($\sim 0.8 \rm M_{\odot}$) and
massive companion ($\sim 11 \rm M_{\odot}$). The subsequent 
lifetime of this companion is $\sim 2 \times 10^8$ years
(approximately the age difference we infer between white dwarf and pulsar), after which it forms a
neutron star. The problem with this scenario is that it will be difficult for the system to
remain bound and with small eccentricity (the post-supernova eccentricity is likely to be
preserved in a system containing compact stars). Although a suitably directed kick velocity
can account for this, the probability of obtaining the correct direction and magnitude
($\sim$ 30 km.s$^{-1}$) is rather small\footnote{In order to maintain small eccentricity, the
pre-supernova orbital period has to be similar to the post-supernova period and thus the kick
velocity is required to match the change in orbital velocity due to mass loss.}. Nevertheless,
if true, this scenario implies a population of isolated pulsars with similar fields and
spin-down times with a binary origin.

 The determination of a spectroscopic gravity for this
star should help to constrain the models, as the helium dwarfs should have $\log~g \sim$ 7
and the carbon dwarfs should have $\log~g \sim$ 7.8.

\subsection{PSR J1713+0747}
 Lundgren et al. (1996b) have measured $V=26.0 \pm$ 0.2 and
$I=24.1 \pm$ 0.1 for this companion. This yields a conservative temperature limit of $<$3800~K.
Camilo, Foster \& Wolszczan (1994), obtained a timing parallax, yielding a distance of 1.2$\pm$0.4 kpc.
 Given this pulsar distance,
 we use the temperature and the absolute magnitude $M_{I}$ = 13.8 $ \pm$ 0.7.
 The characteristic age for this pulsar is 
9.2$\pm$0.4 Gyr, including the correction for proper motion from Camilo et al. 

We can find acceptable solutions for all masses from 0.15-0.31 $\rm M_{\odot}$ with a thick H
envelope and for all masses below 0.27 $\rm M_{\odot}$ for a thin H envelope. However, the 
mass function for this system restricts the companion mass to be $\geq 0.28 \, \rm  M_{\odot}$,
so that many of these solutions are excluded. If we keep only those models with masses greater
than 0.28, then the
cooling age  must lie in the range 6.3-6.8 Gyr. 
We use this result below, but,
note that, using the dispersion measure distance estimates (which allow
smaller distances), we can obtain cooling ages up to 13.2 Gyrs and encompassing the full range
up to 0.45~$M_{\odot}$. 
 If we assume an $n=3$ braking index,
 we may constrain the initial period to lie in the range
$2.18 < P_0 < 2.68$ ms ( the lower limit goes to zero for the dispersion measure distances). 
To be conservative we  also consider the effects of 
 $n=2$. This leads to a more conservative upper limit of 3.0 ms.

Inverting the normal procedure for the mass function, we
can thus constrain the a neutron star mass to be less than 1.64 $\rm M_{\odot}$ (assuming the
parallax distance). 

We can find no consistent C/O solutions for this object.

\subsection{PSR J1640+2224}
Lundgren et al. (1996b) have also detected the companion in this system, with 
$V=26.0 \pm$ 0.3 and $I=24.6 \pm$ 0.2, yielding $ T_{\rm eff} = 4460 \pm 1125$ K.
The characteristic age for this system is at least 16.2 Gyr and the distance 1.2 kpc, so that
(including distance errors) we have an absolute magnitude $M_I = 14.2 \pm$ 0.7.
Once again, the mass function of this system requires the companion to be more massive
than $0.252 \rm M_{\odot}$. 

 We obtain consistent solutions for all masses
0.25-0.45 $ \rm M_{\odot}$. The range of cooling ages obtained is 3.2-12.2 Gyr, 
including both thick and thin hydrogen envelopes. The extremely low $ \dot{P}$ of this 
pulsar means the $ t_P$ is both large and sensitive to $ V_{\perp}$.
 Thus, we can only obtain a lower limit on the initial period
by taking the smallest characteristic age and largest cooling age, yielding $ P_0 > 1.6$
ms.

\subsection{PSR J0034-0534}
The uncorrected characteristic age is 4.4$\pm 0.4$ Gyr and the distance is 1.0 $\pm$ 0.3 kpc. The proper motion
of this pulsar is not known. However, if the pulsar had a transverse velocity of 100 $\rm
km \, s^{-1}$, the characteristic age could be as much as 9.3 Gyr, so we adopt this as an upper limit
to the characteristic age. Lundgren et al. (1996b) 
have detected this companion at $I = 24.8 \pm$ 0.3 and set a limit of $V >$ 26.8, which implies
that the temperature $<$ 3500 K. This is sufficient to constrain the ages of cooling models
of various masses (see Figure~\ref{Solve0034}) to lie within the range 4.4 - 14.9 Gyr.
 However, if we require that the cooling age be less than the 
characteristic age (using our upper limit of 9.3 Gyr), then the mass range is 0.15 - 0.32 $\rm M_{\odot}$ (assuming $n=3$). If we use $n=2$, then all masses are allowed. 
Thus our conservative upper limit on
the initial period is $ P_0 < 1.4$ ms (using $n=2$ and assuming a 100 $\rm km \, s^{-1}$
transverse velocity). A measurement of the proper motion for this
pulsar could improve the constraint on the initial period, although it will not
prove to be a useful constraint on the nuclear equation of state unless the braking index
can be constrained.


\subsection{PSR J2145-0750}

For the carbon/oxygen-core white dwarfs, we use Wood's models (Wood 1992), which have a
more complete description of crystallization,  to perform
the same sort of analysis as above. For PSR J2145-0750, the characteristic age is $>$8.2 Gyr and the
dispersion measure distance is 0.5 kpc. The minimum mass is 0.43 $\rm M_{\odot}$.
Optical observations by Lundgren et al. (1996b) give $V = 23.7\pm$0.1 and $I = 23.0\pm$0.1. The $V-I$
broad band colours of Bergeron et al. (1995) yield a temperature
estimate of 6235 $\pm$ 970 K.

The determination of ages and masses for the carbon sequences requires a little care,
because the cores of $ M > 1 \, \rm M_{\odot}$ white dwarfs can begin to crystallize after $\sim$ 1
Gyr. This means that the absolute magnitude curves for different masses can cross, making the
determination of consistent solutions slightly more complicated. We show the $V$ band
magnitudes for this case because the theoretical $M_{V}$ curves are more dispersed
than the $M_{I}$ curves. Figure~\ref{Solve2145} shows the analysis using the
Wood curves for carbon cores, helium mass fraction = $10^{-2}$ and a hydrogen mass
fraction of $10^{-4}$. For models with pure helium envelopes, the 
age limits are
similar. We find consistent models for all
masses  (in both C and O cases, the cores have begun to
crystallize for the more massive cases), and a range of cooling ages 1.6-6.9 Gyr. 

A transverse velocity of 100 $\rm km \, s^{-1}$ would change the sign of $ \dot{P}$ for this pulsar,
so the characteristic age of 8.2 Gyr is only a lower limit. Using the largest cooling age, we can
thus get a lower limit on the initial period $ P_0 > 6.4$ ms (assuming an $n=3$ braking
index).


\subsection{PSR J1022+10}

This pulsar has a spin period of 16.45 ms and spin down age $>$5.8 Gyr.
 It lies
at a distance of 0.6 kpc and has been observed by Lundgren et al. (1996b), yielding $V = 23.09 \pm$
0.04 and $I = 22.665 \pm$ 0.007. The mass function for this pulsar requires a companion of mass
$ M > 0.72 \, \rm M_{\odot}$, putting this firmly in the carbon/oxygen mass range. Our temperature
estimate for this star is 8050 $\pm$ 950 K.
 Comparison of the observations with the Wood models yields an
age range of 1.5 - 3.9 Gyr and a mass range $ M > 0.7 \, \rm M_{\odot}$. Thus we obtain 
$ P_0 > 9.4$ ms.

\subsection{PSR B0655+64}

This was one of the first pulsars with an optically identified companion. Upper
limits on the proper motion mean that we can constrain the characteristic age to
the range 4.4-146(!) Gyrs. The distance is 0.48 kpc. The mass function constrains
the companion to have $M \geq 0.67 \rm M_{\odot}$.
 van~Kerkwijk \& Kulkarni (1995)
have identified the white dwarf companion as a DQ star, i.e., it shows traces of
molecular carbon in the optical spectrum. This makes accurate temperature determination
difficult but does constrain the temperature to
be in the range 5500-8000 K, when the convection zone is deep enough to dredge
up trace amounts of carbon from the core and deposit them in the atmosphere (Pelletier et al 1988).  Kulkarni
(1986)
has measured $V$ = 22.2 for this star. Since molecular carbon
is seen, the helium and hydrogen envelopes must be quite thin. We use the 
pure helium envelope/carbon core models ($q_{\rm He} = 10^{-4}$) of Wood (1992) to calculate the cooling ages in
Figure~\ref{Solve0655}. We find solutions for all models $\rm M \geq 0.7 \, \rm M_{\odot}$.
The range of cooling ages is 2.5 - 4.7 Gyr.
 The large range in characteristic age means that no useful limits can be obtained for $n=3$.
For $n=2$, the smallest
characteristic age and largest cooling age  yield $\rm P_0 > 94 ms$ (since, in this case, the
maximum cooling age is less than the minimum timing age).


\subsection{PSR 2019+2425}
This pulsar has one of the largest characteristic ages (corrected for proper motion) of any
pulsar, ranging from 14.7 to 112 Gyr, depending on the rather uncertain distance. It lies
close to the galactic plane, which means that extinction is an important problem.
Lundgren et al. (1996b) have detected the pulsar companion at $V=26.4\pm 0.4$ and $I=25.0\pm 0.3$.
To estimate the extinction to the companion, we find the extinction in
the general direction of J2019+2425 from Neckel \& Klare (1980). For distances less than
1 kpc, $ A_V$ rises linearly to 0.4 magnitudes, but then rises steeply to $\sim$ 3
magnitudes
at a distance of 1.5 kpc (i.e. 1.5 magnitudes for $ A_I$).
Using this extinction law and the 30\% distance uncertainty, we find that $V-I$=1.0$\pm$0.4,
which implies $ T_{\rm eff} \sim 4000-6500$ K. Similarly, we estimate the absolute $I$ band 
magnitude as $ M_I$ = 14.9$\pm$1.0. A further constraint is that the mass function requires
that the companion mass $>$ 0.32 $ \rm M_{\odot}$.

Using these limits we find consistent solutions for all masses above the minimum allowed
by the mass function. Figure~\ref{Solve2019} shows the solution for the 0.35 $ \rm
M_{\odot}$ case. The approach taken here is slightly different because the allowed
ranges of $ T_{\rm eff}$ and $ M_I$ are correlated. 
 The large extinction correction implies that
the allowed $V-I$ and $ M_I$ ranges are a function of distance, as shown in
Figure~\ref{Solve2019}. Using this approach we find a range of cooling ages from 7.6-13.9 Gyr.
Although Wood's models again do not extend to low enough temperatures in most cases, modest
extrapolation indicates that consistent solutions may also be found with C/O models for
all masses (a sample 0.6 $\rm M_{\odot}$ model is shown in Figure~\ref{Solve2019}).
This extends the allowed range to 3.4-13.9 Gyr.

Using these values and the range 14.7-112 Gyr for the characteristic age, we may derive a lower
limit to the
 initial period of 0.9 ms for $n=3$ (or 2.1 ms for $n=2$). Despite the uncertainty
in the cooling age resulting from the extinction correction, we find that this result is
quite robust, primarily because the measured proper motion constrains the characteristic age to
be significantly greater than the age of the galaxy.


\subsection{PSRs B1855+09 and J0751+1807}
For those systems where we have only upper limits on the companion magnitudes, it
 is
sometimes still possible to place some constraints on the ages and masses of the
 stars
involved.

For PSR 1855+09, the almost 90$^{\circ}$ inclination of this orbit has allowed 
Kaspi et al. (1994)
to determine the masses of both components as well as the parallax and distance.
 The companion mass
is $0.258^{+0.028}_{-0.016} \rm  M_{\odot}$ and the distance is $0.9^{+0.4}_{-0.2
}$ kpc.
 The
characteristic age, corrected for kinematic effects, is $4.95\pm0.05 \times
 10^9$ years. We can
thus restrict our model comparisons to the 0.25 $\rm M_{\odot}$ model in this 
case.
Kulkarni, Djorgovski \& Klemola (1991) obtained limits of $R>$ 24.6 and $I>$ 23.4, while
 Callanan et 
al (1989) obtained
$V>$ 25.4. Callanan et al. estimated the extinction to the system to be $ A_V
= 1.5-2.0$
mags kpc$^{-1}$. Using the extinction law of Savage and Mathis (1979), this becomes $
 A_R$=1.1-1.5 and
$ A_I$=0.7-1.0 mags kpc$^{-1}$. Figure~\ref{SolveVIT} shows the comparison of the 
model curves with
each constraint. We use two constraints, using the largest extinctions and furthest
 distance
estimate; and the smallest extinctions and closest distance estimates, to examine
 the range of
possible solutions. We see that the 0.25 $\rm M_{\odot}$ sequence is perfectly 
consistent with
the bounds on the absolute magnitude for all ages between the characteristic age of 4.95
 Gyr to a
minimum age of 1.4 Gyr. This yields a constraint $ P_0 > 4.5$ ms.

Also shown in Figure~\ref{SolveVIT} are the detection limits for the companion
to J0751+1807,
 with a characteristic age of 6.3-12.2 Gyr.
The only meaningful limits that can be placed
on this system is that the cooling time is longer than 0.8 Gyrs. So, this could still
be a slightly older version of the J1012+5307 system. Given the small orbital
period we might expect a low companion mass ($\sim 0.15 \rm M_{\odot}$) so that
extending the detection limit to $V\sim 26$ would detect the companion if it were
younger than 2~Gyrs. Such a detection would be extremely interesting.


The various constraints on the cooling ages are collected together in Figure~\ref{tCool}.


\section{Discussion}
\label{Discuss}

\subsection{Binary Evolution}

The formation of low mass binary pulsars (LMBPs) has been discussed by many authors 
(for reviews see Phinney and Kulkarni 
1994; Verbunt 1993 and references therein). If we consider a binary containing a pulsar
and a stellar companion,
  the binary will undergo mass transfer if the non-degenerate companion
begins to expand as a result of nuclear evolution or if the orbit decreases due to magnetic
braking or gravitational wave radiation. This mass transfer results in the spin-up of
the neutron star to form a millisecond pulsar. The mass loss also means that the companion never
evolves far enough to grow a core of mass large enough to ignite helium. Rather, the envelope is
lost during the course of the evolution and the remnant of the secondary settles down to a
degenerate configuration, a low mass helium core white dwarf. The fact that giants have a well-defined
relationship between core mass and giant radius, allied with the fact that the star must fill
its Roche lobe to lose matter to the companion (assuming corotation), means that there exists
a relationship between orbital period and secondary mass in the LMBPs
(Refsdal \& Weigart 1971; 
Joss, Rappaport \& Lewis 1983;
Rappaport et al. 1993). However, this holds only as long as the secondary star is sufficiently evolved to
have a convective envelope when it overflows its Roche lobe. This is because mass loss from the
secondary results in expansion of the orbit, shutting off mass loss unless the donor star increases
as well (which requires a convective envelope, rather than a radiative one).
 Thus, this scenario describes systems with orbital periods $\sim 50 - 10^3$ days.

For shorter period systems, the donor star overflows its Roche lobe either on the main sequence
or during the transition from the
main sequence to the giant branch. The envelope is still primarily radiative in this case, and the
star will shrink in response to mass loss. The result is that one needs angular momentum loss
mechanisms such as gravitational radiation and magnetic braking to maintain mass transfer in
these systems. The competition between these loss mechanisms and the mass-transfer induced
evolution of the system leads to a very steep relationship between 
final orbital period and initial orbital period/final core mass 
(Pylyser and Savonije 1988; Cot$\rm \acute{e}$ and Pylyser
1989). 

In Figure~\ref{mcp} we compare our mass determinations for these companions with the results of
Rappaport et al. (1993) and Pylyser \& Savonije (1988) (omitting models in the latter sample where
the accretor was far from 1 $\rm M_{\odot}$). We find excellent agreement with the 
models, especially that of Pylyser \& Savonije, where the mass estimates of B1855+09, J0034-0534,
 J0437-4715 and
J1012+5307 trace out the mass-period relation very nicely. 
Thus the two different scenarios provide a natural explanation for
the orbital period gap between 20 and 50 days.


\subsection{Neutron Star Spin-up}

Under the assumption that the magnetic field does not decay, the cooling age of the white
dwarf allows us to estimate the initial spin period of the millisecond pulsar, by inverting
formula (\ref{spin-down}). These estimates are shown in Table~\ref{P0}.

The simplest theories regarding spin-up of recycled pulsars to millisecond
periods (Smarr \& Blandford 1976; Ghosh 1995 and references therein) predict that the initial period
should be equal to the equilibrium spin-period of the neutron star of 
magnetic field $B$ accreting at a rate $ \dot{M}$. This predicts an initial
spin period 
\be P_0 = 1.89\, {\rm ms} \, B_9^{6/7} \left( \frac{\dot{M}}{\dot{M}_{\rm Edd}} \right)^{-3/7}, \label{peq} \ee
where $ \dot{M}_{\rm Edd}$ is the Eddington accretion rate. Thus, a comparison between the inferred initial spin
period and magnetic field can determine the accretion rate onto the neutron star during spin-up.
However, this inversion is complicated somewhat by the uncertainty in the macroscopic dimensions
of the neutron star. Figure~\ref{bp0} shows the spin period-magnetic field diagram for the
millisecond pulsars, scaled in such a way that it reflects observational parameters ($P$ and $\dot{P}$) only. This is done by plotting
\be \frac{\mu_{26}^2}{I_{45}} = 1.026 \left(\frac{P}{1 \, \rm ms}\right) \left(\frac{\dot{P}}{10^{-20}}
\right) \ee
where $\mu = R^3 B$ is the magnetic moment (in this case expressed in terms of units of $10^{26} \rm
G cm^{3}$) and $I$ is the moment of inertia of the pulsar.
Thus we infer accretion rates
$\sim 10^{-2} - 0.1 \, \dot{M}_{\rm Edd}$, possibly
even somewhat lower in some cases.
A similar conclusion was reached by Lundgren et al. (1996).
 This is consistent with the findings that low mass
X-ray binaries have a range of sub-Eddington accretion rates (Bradt \& McClintock 1983).


\subsection{Magnetic Field Decay}

Although many authors have noted that millisecond pulsars must have very long magnetic field
decay times (Kulkarni 1986; Camilo et al. 1994), the determination of a cooling age allows us to make a quantitative estimate of
$ t_{\rm D}$ in the context of the paradigm outlined in section~\ref{Pulsar}. To get the lower limits on $
t_{\rm D}$ using equation~(\ref{decay2}), we
must use the lower limits on both the distance and the cooling age (and assume a value for n; we will use
n=3 below). 
For pulsar J1713+0747 we find a lower limit of 15.6 Gyrs for $t_D$ (essentially because the cooling
age is restricted to be quite close to the spin-down age), which is the largest value amongst all
our sample. PSR~J0034-0534 also yields a strong constraint (6.5 Gyr) although most of the other
systems yield weaker constraints ($\sim 1$~Gyr in most cases). These results strongly support
the view that millisecond pulsar fields do not decay at all. Figure~\ref{tD} shows the limits using
equation~(\ref{decay2}) 
for the various binary systems.


In equation~(\ref{decay2}) we  neglect $\rm P_0$ to obtain a lower limit on $t_D$.
However, if we assume that the
pulsar did begin  the millisecond pulsar stage with $ P_0 \ll P$, then we may also
obtain  an upper limit on $ t_{\rm D}$ by using the maximum allowed cooling age and minimum $t_P$.
For most of the pulsars here this is not an interesting limit, but for J1012+5307, we find $ t_{\rm D} < 0.2$ Gyr. Thus, we have an alternative explanation for this pulsar's anomalous parameters. This pulsar was
either born
spinning close to it's current period or its magnetic field decays on a timescale $\sim 10^8$
 years.

\subsection{Nuclear Equation of State}

In addition to constraining evolution scenarios, we note that our models offer the possibility of
determining neutron star masses and hence constraining the nuclear equation of state.
If we adopt the van Kerkwijk et al. gravity in  
 section~\ref{J1012}, then the neutron star mass is $> 1.7 M_{\odot}$. This would rule out 5 of the
softer equations of state listed in
 Cook et al. (1995). This is also interesting, because it raises the value of the minimum rotation
period for a neutron star. Of the surviving models, the shortest minimum spin period is 0.47 ms.
However, good gravity measurements and good radial velocities are required to make this a
robust calculation.

If one assumes $n=3$, the initial spin period of J0034-0534 can also provide a useful
constraint on the harder equations of state,
 but this conclusion is not robust, because a
braking index $n=2$ will yield an upper limit that lies above all minimum spin periods.

In conclusion, we have shown that using the white dwarf cooling ages as an independent chronometer for 
binary pulsars can teach us a lot about both 
 neutron star structure and binary evolution. In
particular, the determination of initial spin periods provides us with new information about the
final stages of pulsar spin-up and evolution. However, we must reiterate that our results should
be considered as preliminary until certain uncertainties can be conclusively addressed.
 In particular, except in the few cases where parallaxes are available, we are forced to use
dispersion measure distances. The oft-claimed 30~\% error is a statistical statement; individual
cases may be more uncertain, as is suggested by the case of PSR B0820+02. Nevertheless, the good
agreement we obtain argues that this is a reasonable method in general. Secondly, our temperature
estimates are uncertain for some of the cooler members of our sample. While we have used the
results of Bergeron et al to constrain these errors, it should be possible to obtain similarly
accurate atmospheres for these lower-gravity cases, which will conclusively settle the issue.
Finally we look forward to
 further observations of low mass binary pulsars,
both in radio and optical, which will undoubtedly lead to even better constraints in the future.

The authors would like to thank Marten van Kerwijk for use of 
results prior to publication and discussion of white dwarf observational uncertainties
and Glenn Soberman for discussions about mass transfer in binaries, as well as the
referee, Frank Verbunt, whose comments contributed greatly to the
clarity of this paper. This work was supported by
NSF grant AST93-15455 and NASA grant NAG5-2756.

\label{lastpage}
\clearpage

\begin{table*}
\centering
\begin{minipage}{140mm}
\caption{Timing Information for Optically Measured Pulsar Binaries \label{TimID}}
\begin{centering}
\begin{tabular}{lclcrlccc}
Name & $P$ & $\dot{P}$ & f(M)& $P_{\rm orb}$ & D & $\mu$ & $t_p$ & refs \\
 & (ms) & $10^{-20}$ & ($10^{-3} M_{\odot}$)& (days) & (kpc) & (mas.yr$^{-1})$& (Gyr) & \\
\hline
\multicolumn{9}{c}{Helium Cores} \\
J0034-0534 & 1.877 & 0.67(6) & 1.260 & 1.589 & 1.0 & \cd & 4.1-9.3 & 1 \\
J1713+0747 & 4.570 & 0.853(2) & 7.872 & 67.825 & 1.2(4) & 6.4(8) & 9.2(4) & 2 \\
J0437-4715 & 5.757 & 5.71(1) & 1.239 & 5.741 & 0.178(26) & 135(4) & 4.4-49.1 & 3,4 \\
J1640+2224 & 3.163 & 0.29(2) & 5.889 & 175.461 & 1.2 & \cd & $>$ 16.2 & 5 \\
J1012+5307 & 5.255 & 1.46(8) & 0.577 & 0.605 & 0.5 & \cd & $>$ 5.4 & 6 \\
B0820+02   & 864.8 & 10390(30) & 3.004 & 1232.5 & 1.4 & $<$ 13.6 & 0.13 & 7 \\
B1855+09   & 5.362 & 1.784 & 5.540 & 12.327 & 1.0(3) & 6.16(8) & 4.95(5) & 8 \\
J0751+1807 & 3.479 & 0.80(8) & 0.964 & 0.263 & 2.0 & \cd & 6.3-12.2 & 9 \\
J2019+2425 & 3.935 & 0.702(2) & 10.65 & 76.512 & 0.9 & 23(1) & 14.7-112 & 10 \\
\multicolumn{9}{c}{Carbon Cores} \\
J2145-0750 & 16.05 & 2.9(2) & 24.03 & 6.839 & 0.5 & \cd & $>$8.2 & 1 \\
J1022+1001 & 16.45 & 4.2(3) & 82.79 & 7.805 & 0.6 & \cd & $>$5.8 & 11 \\
B0655+64   & 195.7 & 69(2) & 70.95 & 1.029 & 0.5 & $<$46 & 4.4-146 & 12 \\
\hline
\end{tabular}
\end{centering}
{\bf References:} (1) Bailes et al. (1994); (2) Foster et al. (1993); (3) Johnston et al.
(1993); (4) Bell et al. (1995); (5) Foster et al. (1995); (6) Nicastro et al. (1995)
(7) Taylor \& Dewey (1988);
(8) Kaspi, Taylor \& Ryba (1994) ; (9) Lundgren et al (1995); 
(10) Nice et al. (1993); (11) Camilo (1996); (12) Jones \& Lyne (1988)

\vspace{5mm}

All distances without quoted errors are assumed to be 
subject to a 30\% distance error due to the electron density model of Taylor \& Cordes.

\end{minipage}
\end{table*}

\begin{table*}
\centering
\begin{minipage}{140mm}
\caption{Optically Identified White Dwarf Companions to Pulsars \label{OptID}}
\begin{centering}
\begin{tabular}{lccccccc}
Name & E$_{B-V}$ 
 & $T_{\rm eff}$
 & $m_{B}$ & m$_{V}$ & m$_{
R}$ & m$_{ I}$ & refs
 \\
 & & (K) & & & & & \\
\hline
\multicolumn{7}{c}{Helium Cores}\\
J0034-0534& 0.00 &  $<$ 3500 & \cd & $>$ 26.8 &$>$ 25.0 & 24.8(3) & 1,2 \\
J1713+0747 & 0.08 &  3430(270) & $>$ 27.1 & 26.0(2) & \cd &
24.1(1) & 1 \\
J0437-4715 & 0.07 &  4610(200) & 22.19(8) & 20.84(2) & 20.07(3) & 19.51(5) & 3,4,5 \\
J1640+2224 & 0.05 &  4460(1125) & \cd & 26.0(3) & 24.5(3) & 24.6(2)
 & 1,6 \\
J1012+5307 & 0.00 &  8550(25)\footnote{
The error is much smaller than the others because this temperature was determined
spectroscopically, rather than photometrically.} & 19.78(4) & 19.58(2)
 & 19.49(4) & 19.32(4) & 7, 8 \\
B0820+02 & 0.03 &  15250(1250) & \cd & 22.8(1) & \cd & \cd & 9,10 \\
B1855+09 & 0.5 &  \cd & \cd & $>$ 25.4 & $>24.6$ & $>23.4$ & 11,12 \\
J0751+1807 & 0.03 & \cd & \cd & $>$23.5 & \cd & \cd & 8 \\
J2019+2425 & $>0.2$ & 5400(1300) & \cd & 26.4(4) & \cd & 25.0(3) & 6 \\
\multicolumn{7}{c}{Carbon/Oxygen Cores} \\
J2145-0750 & 0.03 &  6210(890) & 23.89(11) & 23.7(1) & \cd &
22.97(7) & 1 \\
J1022+1001 & 0.00 &  8050(950) & \cd & 23.10(4) & \cd & 22.665(7)
 & 1 \\
B0655+64 & 0.05 &  7500(1500) & \cd & 22.2 & 22.1 & \cd & 13 \\ \hline
\end{tabular}
\end{centering}

{\bf References:} (1) Lundgren et al. (1996b); (2) Bell et al. (1995); (3) Danziger et al. (1993);
 (4) Bell et al. (1993); (5) Bailyn (1993); (6) Lundgren at al (1996a); (7) van~Kerkwijk et
al (1996); (8) Lorimer
et al. (1995a); (9) van~Kerkwijk \& Kulkarni (1995); (10) Koester et al. (1992);
(11) Kulkarni et al. (1991); (12) Callanan et al. (1989); (13) Kulkarni (1986).
\end{minipage}
\end{table*}

\begin{table*}
\centering
\begin{minipage}{140mm}
\caption{Cooling Ages and Initial Spin Periods \label{P0}}
\begin{centering}
\begin{tabular}{lcccccc}
Name & $P$  & $t_{P}$
 & $t_{\rm cool}$ & $P_0$\footnote{Based on $n=3$}  & $P_0$
\footnote{Based on $n=2$}& $t_{\rm D}$ \\
 & (ms) & (Gyr) & (Gyr) &
(ms)
 & (ms) & Gyr \\
\hline
\multicolumn{7}{c}{Helium Cores} \\
J0034-0534 & 1.88 & 4.1-9.3 & 4.4-14.9 & $<$ 1.4 & $<$ 1.4 & $>$6.5 \\
J1713+0747 & 4.57  & 8.8-9.6 & 6.3-6.8 & 2.2-2.7 & 2.8-3.1 & $>$15.6
\\
J0437-4715 & 5.76 & 4.5-50.3 & 2.5-5.3 & $<$ 5.3
 & 2.4-5.3 & $>$ 1.1
 \\
J1640+2224 & 3.16 & $>$16.2 & 3.2-12.2 & $>$ 1.6 & $>$ 2.0 & $<$ 45
 \\
J1012+5307 & 5.26 & $>$5.4  & $<$ 0.4 & $>$ 5.1 & $>$ 5.1 & $<$0.2
 \\
B1855+09 & 5.36 & 4.9-5.0 & $>$ 1.4 & $<$ 4.5 & $<$ 4.6 & $>$1.3\\
J0751+1807 & 3.48 & $>$6.3 & $>$0.8 & $\cdots$ & $\cdots$ & $>$0.4\\
J2019+2425 & 3.94 & 14.7-112 & 3.4-14 & 0.9-3.9  & 2.1-3.9 & $>$1.3\\
\multicolumn{6}{c}{Carbon/Oxygen Cores} \\
J2145-0250 & 16.05 & $>$ 8.2 & 1.6-6.9  &
 $>$ 6.4  & $>$ 9.3 & $<$ 39.7\\
J1022+1001 & 16.45 & $>$5.8 & 1.5-3.9 & $>$9.4 & $>$10.9 & $<$ 10.5 \\
B0655+64 & 195.7 & 4.4-202 & 2.5-4.7 & $\cdots$  & $>$91  & $>$0.9 \\
\hline
\end{tabular}
\end{centering}
\end{minipage}
\end{table*}

\clearpage
\setcounter{figure}{0}

\begin{figure}
\caption{The heavy solid line indicates the black body relation (BB) between $T_{\rm eff}$ and
$V-I$. The thin solid curve labelled BSW indicate the colours from the stellar atmosphere
calculations of Bergeron, Saumon \& Wesemael (1995) for
$\log g=7.5$. The left and right panels show two different colour ranges. In the right panel the
shaded area shows the range in which the true colours are expected to lie. Note that we have had
to extrapolate the BSW values to temperature below 4000~K. This turns out to be unimportant, since we are only able to derive upper
limits on the temperatures for the two systems in this region anyway.
The colours and corresponding temperature estimates are
shown for various individual systems in solid circles. Pulsar J1640+2224 is omitted because the
large error bars span both panels, but the temperature inference procedure is the same.
\label{tc} }
\end{figure}

\begin{figure}
\caption{{\bf J0437-4715:}
We show the effective temperature and absolute I magnitude as a function of age for
cooling sequences of mass 0.15 to 0.45 $\rm M_{\odot}$, in steps of 0.05 $ \rm M_{\odot}$.
The heavy solid line is the 0.20 $\rm M_{\odot}$ model.
 The
horizontal dotted
lines in the upper panel indicate the allowed range of $ T_{\rm eff}$. In the lower panel, the
dotted lines indicate the allowed range in absolute magnitude from the observations using the
timing parallax distance (the upper limit of the absolute magnitude due to the Bell et al
measurement is also shown).
In order for a solution to be consistent, it must satisfy both observational criteria.
The vertical shaded regions indicate two consistent 
solutions. Given the apparent I magnitude, each absolute
magnitude corresponds to a different distance, and hence a different characteristic age (once
corrected for the Shklovskii term). This is shown by
the heavy dashed lines and the lightly shaded region, corresponding to a spin-down index of $n=3$.
The width of this region is dominated by the  uncertainty in the proper motion.
The
models shown here are for the `thick' hydrogen layer models (as defined in paper~I). The horizontal bar indicates the allowed age range.
\label{Solve0437}}
\end{figure}

\begin{figure}
\caption{The upper panel demonstrates the constraints on the relationship between neutron star
and companion mass. The heavy solid line is the upper limit on $M_{ns}$ obtained from the mass
function and the Sandhu et al limit on the inclination. The dotted lines indicate the same relation
for other allowed inclinations. The solid points in the lower panel span the distance range allowed
for each white dwarf mass given the temperature constraints. The horizontal dashed lines are the
timing parallax distance limits and the vertical dashed line at 0.22 $M_{\odot}$ is obtained from the
requirement that $M_{ns}>1.4 M_{\odot}$. Thus, the heavy shaded region is that range of distance and
companion mass allowed by all contraints. The lightly shaded region covers the range where the n=3
timing age is expected to lie (the uncertainty in the proper motion dominates).
\label{md}}
\end{figure}

\begin{figure}
\caption{ {\bf J1012+5307:}
 Here we use the effective temperature, luminosity and gravity to
constrain the age and mass of the white dwarf companion. The solid lines
are models with a thick hydrogen atmosphere and masses 0.15, 0.17 and 0.20
$ \rm M_{\odot}$ respectively. The two horizontal shaded regions are the 
gravity values inferred by van Kerwijk et al (1996) and Koester \& Callanan (1997).
The vertical shaded region shows a consistent solution for the
0.2 thick H model and the van Kerkwijk et al gravity, which yields an age significantly
younger than the spin down age, shown by the vertical dotted line at the
far right. The Koester \& Callanan gravity will provide a consistent solution for the
0.15 $M_{\odot}$ model.
\label{Solve1012}}
\end{figure}

\begin{figure}
\caption{{\bf Equation of State Constraints from J1012+5307:}
 The heavy solid line represents the limit from the mass function $f(M)$ measured
from the pulsar timing. The diagonal dotted line is the original mass ratio of
van Kerkwijk et al, and the solid line the revised value.
The two error bars at the bottom indicate the companion masses inferred from the
spectroscopic gravity measurements of Koester \& Callanan (A) and van Kerkwijk et
al (B). The corresponding neutron star mass estimates are given by the error bars
on the right. The horizontal dashed line indicates the standard chandrasekhar value
for the neutron star mass. It is consistent with the A estimate, although the B estimate
requires a value $\sim 0.3 M_{\odot}$ larger.
\label{mass}}
\end{figure}

\begin{figure}
\caption{ {\bf B0820+02:}
 The dashed curves are the models for helium core white dwarfs, while the solid curves are
for carbon core white dwarfs. The vertical dotted lines indicates the characteristic age of the pulsar
for $n=3$ (lower value) and $n=2$.
The solution for the 0.8 $\rm M_{\odot}$ model is shown as a heavy solid
line. The age range consistent with the temperature and dispersion measure constraints
is shown by the shaded region. \label{Solve0820}}
\end{figure}

\begin{figure}
\caption{ {\bf J0034-0534:} The shaded regions indicate consistent cooling solutions for
0.15, 0.25, 0.35 and 0.45 $\rm M_{\odot}$.
The size of the horizontal
arrow indicates the change in $t_{P}$ that would result from a transverse velocity
of 100 $\rm km \, s^{-1}$. \label{Solve0034}}
\end{figure}

\begin{figure}
\caption{ {\bf J2145-0750:} The solid lines indicate carbon core sequences of mass 0.4, 0.7, 0
.8 and
1.0 $\rm M_{\odot}$ respectively. 
The most massive models begin to crystallize after 0.8 Gyr, and the least
massive after 3.6 Gyr. The shaded regions show the consistent solutions for the
0.4 and 0.8
$\rm M_{\odot}$ model. The other solutions are omitted because the crossing of the
model curves means they lie largely on top of one another.
\label{Solve2145}}
\end{figure}

\begin{figure}
\caption{ {\bf B0655+64:}
We show here the curves for 0.7, 0.9 and 1.0 $\rm M_{\odot}$ and a helium envelope
mass fraction = $10^{-4}$. These models have no hydrogen envelope. The shaded region
shows the solution for 0.7 $\rm M_{\odot}$.
\label{Solve0655}}
\end{figure}

\begin{figure}
\caption{{\bf J2019+2425:}
The solid lines are the models for 0.35, 0.40 and 0.45 $\rm M_{\odot}$. The thin
dashed line is for a 0.30 $\rm M_{\odot}$ He core model and
the heavy dashed line is for a 0.60 $\rm M_{\odot}$ carbon core model.
 The dotted line in the upper panel indicates
the allowed range of $V-I$ for each $ M_I$ (and hence each distance). The shaded region
demonstrates the consistent solution for the 0.35 $\rm M_{\odot}$ model. The filled
circle (with error bars) is the allowed range at the nominal dispersion measure distance.
In the bottom panel the vertical and horizontal bars indicate the allowed range of
cooling ages (for all models) and the expected range in $ M_I$ consistent with the
dispersion measure uncertainty. \label{Solve2019}}
\end{figure}

\begin{figure}
\caption{{\bf Limits for Other Pulsar Companions: }
 We compare the absolute V magnitudes of our models with the
limits determined for two other binary pulsar systems. The vertical dotted lines indicate
the characteristic ages for each binary system and the horizontal dotted lines are the
magnitude limits determined from the observations. The models are for
 0.15, 0.25, 0.35 and 0.45
$ \rm M_{\odot}$ respectively. The heavy solid line indicates the 0.25 $\rm M_{\odot}$
model (for comparison with the B1855+09 limits). \label{SolveVIT}}
\end{figure}

\begin{figure}
\caption{ {\bf Cooling Ages for Millisecond Pulsar Companions:}
Here we show the constraints on the cooling age for the various binaries
discussed in this paper. The open circles indicate characteristic ages (for a braking index $n=3$),
 and the filled circles
indicate cooling ages.
The uncertainties in the characteristic ages are because of the Shklovskii term.
The systems are separated into C/O and He white dwarfs (B0820+02
is placed in limbo due to the uncertainty in its mass) and ordered in increasing orbital
period. The vertical dotted line indicates an age of 15 Gyr. \label{tCool}}
\end{figure}

\begin{figure}
\caption{ {\bf The Orbital Period-Mass Relation:}
The upper shaded region is the parameter space spanned by the models of
Rappaport et al. (1993) while the lower shaded region is that spanned by the models
of  Pylyser \& Savonije (1988). Also shown are the mass constraints for various systems
determined in this paper as well as those of Kaspi et al. (1993) for B1855+09 and
the mass function limits for those systems without further constraints. The dotted
line at the bottom of the diagram indicates the boundary below which the companion will
spiral into a 1.4 $\rm M_{\odot}$ neutron star in a Hubble time due to gravitational radiation.
The vertical dotted line indicates the dividing line between carbon cores ($> 0.5 \rm
M_{\odot}$) and helium cores ($< 0.5 \rm M_{\odot}$).
The error bars for B0820+02 and J2019+2425 are from two overlapping mass estimates,
 one for a helium core white dwarf and one for a carbon core.
  J0034-0534 has two upper mass limits, for $n=3$ (0.32 $\rm M_{\odot}$) and $n=2$ (
0.45 $\rm M_{\odot}$). \label{mcp}}
\end{figure}

\begin{figure}
\caption{ {\bf Initial spin periods for millisecond pulsars: } We show here the inferred limits
 on the initial spin periods for the neutron stars
discussed in the text. The three parallel dotted lines indicate the spin-up lines
(\protect{\ref{peq}}) for $ \dot{M}/\dot{M}_{Edd}$ = 1, 0.1 and 0.01 respectively, and
canonical neutron star values of $M$ = 1.4 $\rm M_{\odot}$ and $R$ = 10 km. The heavy dashed lines
indicate the Eddington rate spin up lines for different equations of state taken from
 Cook et al.
(1995), and spanning the range from hard (L) to soft (AU). 
The shaded region represent the case for maximal spin-up 
covered by the range of equations of state (i.e., they result
from the spin-up of a maximum static mass model).
Many of the pulsar magnetic fields are uncertain because of the Shklovski effect on $ \dot{P}$.
The derivation of the limits on $ P_0$ are discussed in the text. The open circles indicate
systems with orbital periods $> 50$ days and closed circles indicate orbital periods $< 50$ days.
The two starred systems are spiral-in systems containing carbon core companions.
\label{bp0}}
\end{figure}
                                                                                            
\begin{figure}
\caption{ The heavy solid curve describes the universal function $ (e^x-1)/x$ where
$x = 2 t/t_{\rm D}$ and $t$, $t_{\rm D}$ are the cooling time and magnetic field decay
time respectively. The dotted lines are the upper limits placed for the various systems
as labelled. The dashed line shows the lower limit on x derived for J1012+5307. 
\label{tD}}
\end{figure}

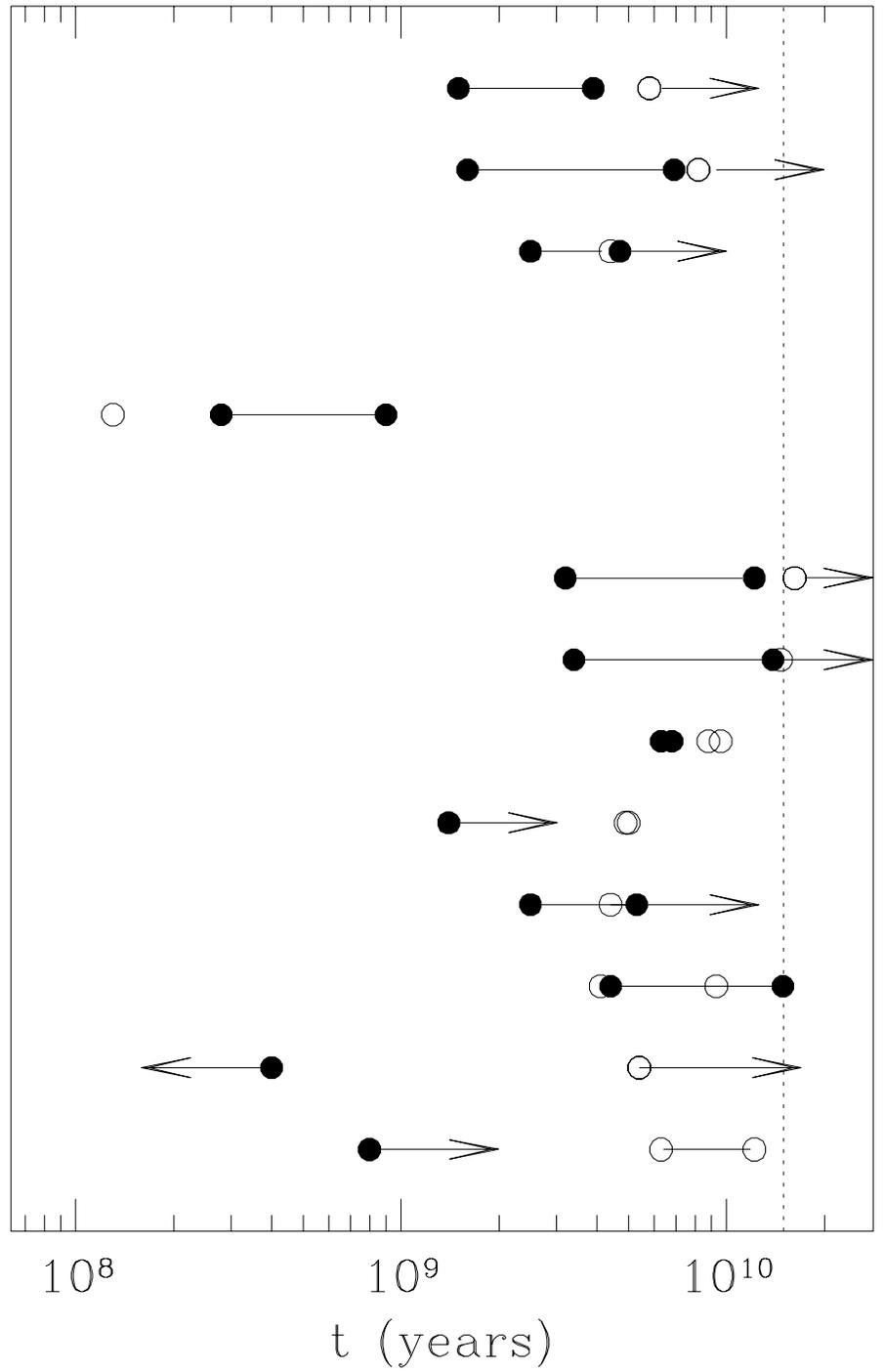
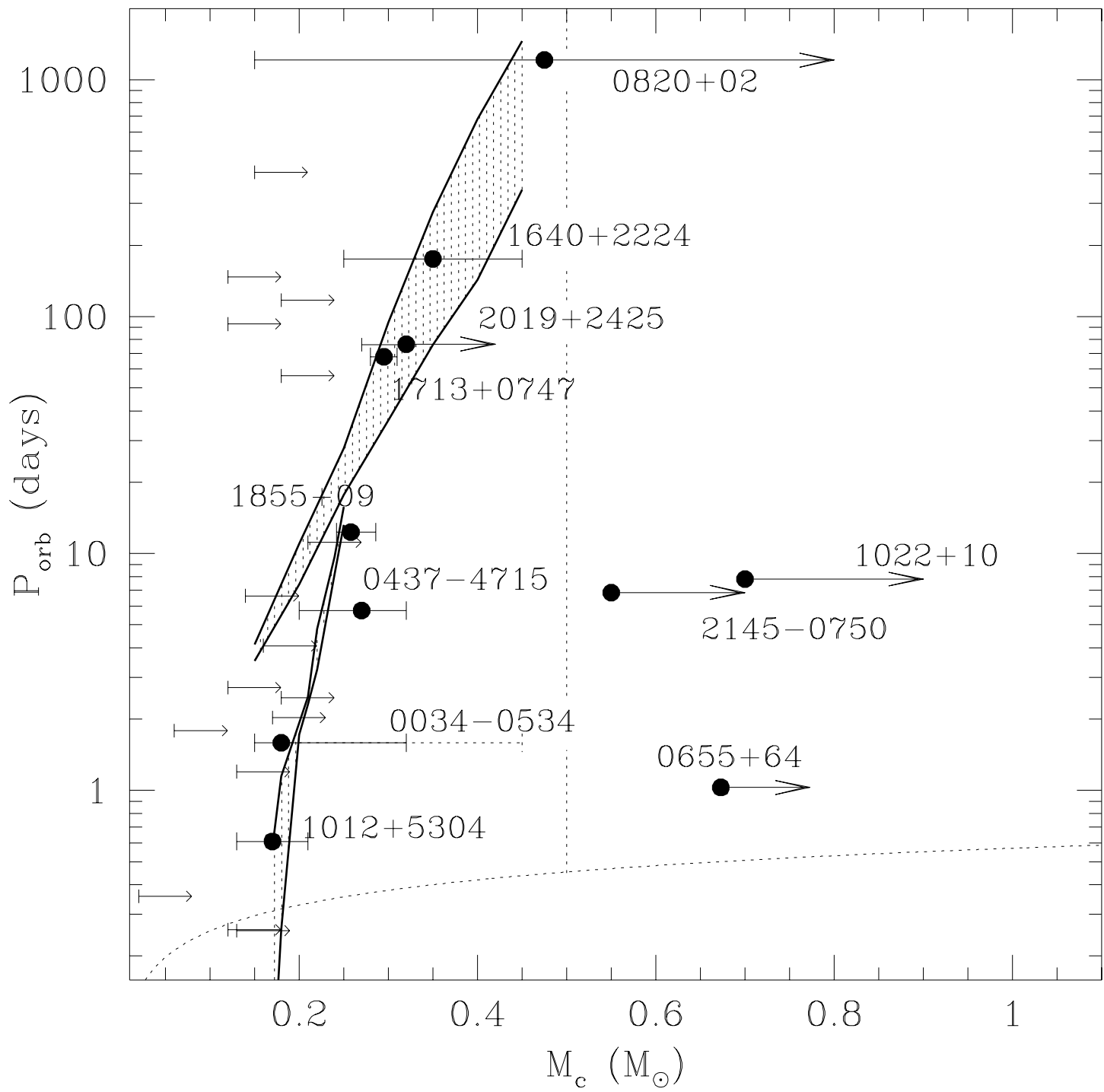
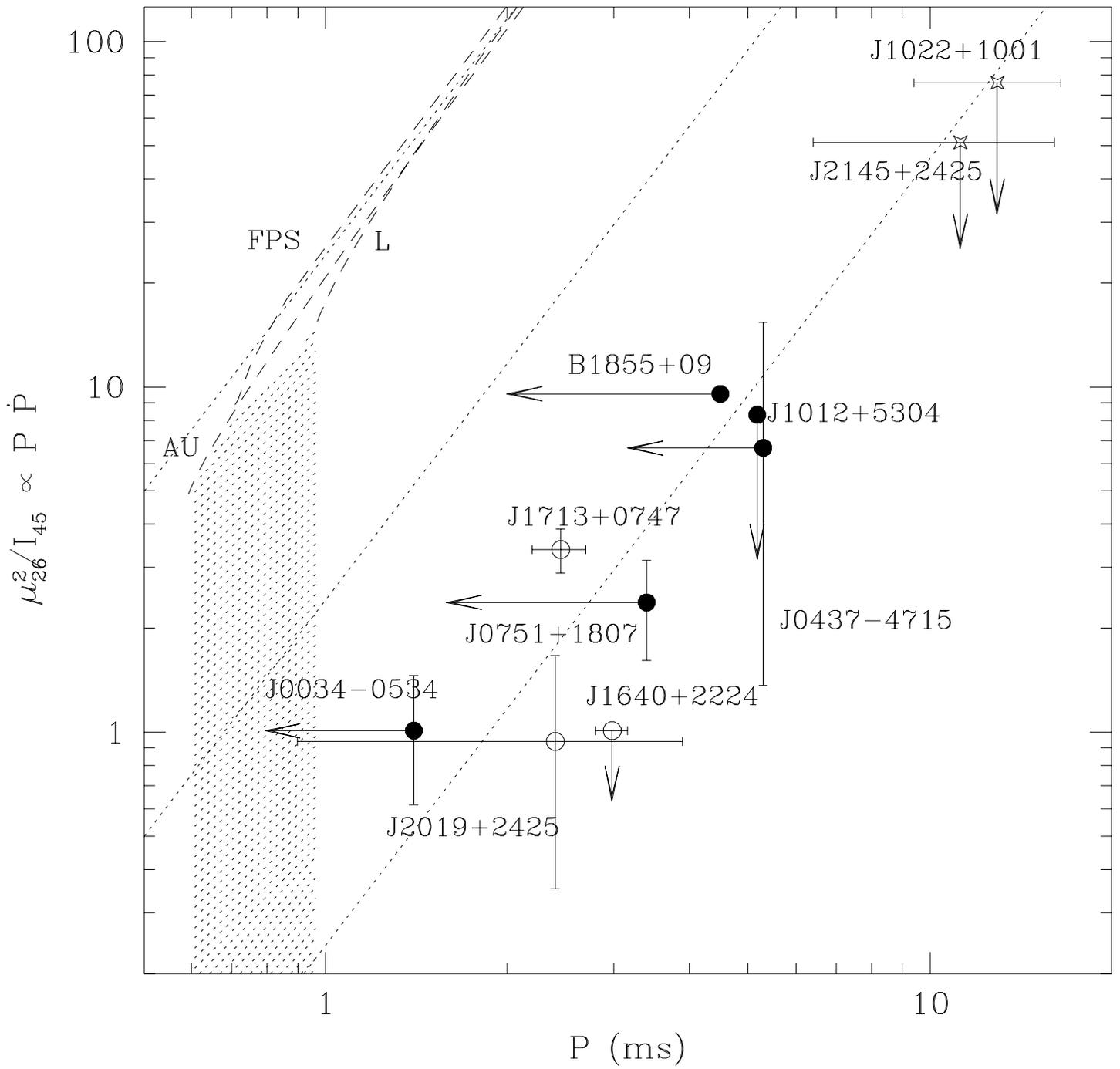
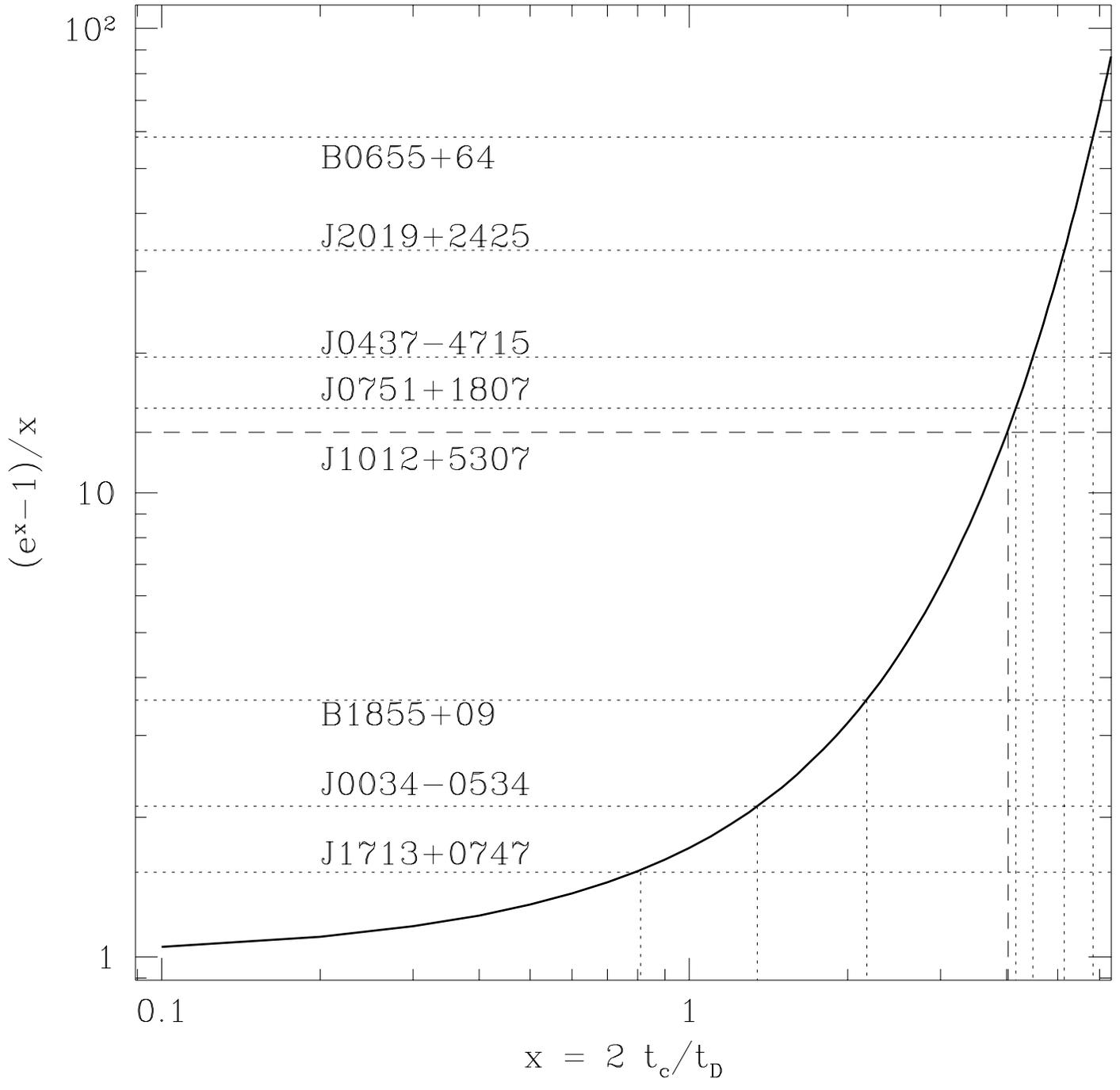
\end{document}